
\font\titlefont = cmr10 scaled \magstep2
\magnification=\magstep1
\vsize=20truecm
\voffset=1.75truecm
\hsize=14truecm
\hoffset=1.75truecm
\baselineskip=20pt

\settabs 18 \columns

\def\b{\bigskip}
\def\bb{\bigskip\bigskip}

\def\ce{\centerline}

\def\no{\noindent}




 \rightline{ UMDHEP 93-199}
\rightline{ May 1993}
\bb

\b
\ce{\titlefont{Effect of CP-Odd Planck Scale Induced Operators}}
\ce{\titlefont{ on the Majoron Scale
    \footnote\dag{\rm{ Work supported by a grant from the National
          Science Foundation}} }}
\bb
\ce{\bf{R.N. Mohapatra and X. Zhang
   }}

\ce{\it{ Department of Physics and Astronomy}}
\ce{\it{University of Maryland}}
\ce{\it{ College Park, MD 20742 }}
\b
\ce{\bf Abstract}

\no
We extend a recent work on the effect of non-perturbative Planck
scale effects on models with spontaneously broken global B-L symmetry,
where it was shown that scale $V_{BL}$ of this symmetry must
be less than 10 TeV. We show that, if the Planck scale effects break
CP invariance, an allowed window for $V_{BL}$ appears where
$10^{8}{\rm GeV}\leq V_{BL} \leq 10^{11}{\rm GeV}$.

 \filbreak
In recent months, there has been growing interest in the possibility[1]
that non-perturbative gravitational effects may not respect any global
 symmetries of nature. Early work on ``no hair" theorem[2] and
recent analyses of the effect of ``worm-hole" solutions on low energy
particle physics seem
to provide evidence for this speculation. In this brief report, we
comment on the impact of this speculation on one particular class
of extensions of the standard model where small neutrino mass arises
in a natural manner, the singlet majoron model[3].

In a recent analysis[4] of the Planck scale effects on the singlet Majoron
model, it was concluded that, there must be an upper bound on the
$V_{BL}$, the scale of B-L symmetry breaking {\it i.e.}

$${
  V_{BL} < {( {m_\nu \over 25 {\rm eV}} )}^{4/7} \times 10~TeV .
  }\eqno(1)$$
\no The basic idea is that, the Planck scale effects are expected in general
to induce non-renormalizable terms of the following
type in the low energy effective lagrangian:

$$
{\eqalign{ V_{Planck}= & {1\over M_{Planck}}
           ( \lambda_1 \sigma^5 + \lambda_2 \sigma^4 \sigma^{*}
+ \lambda_3 \sigma^3 \sigma^{*^2} ) \cr
                       & + {1\over M_{Planck}}( \lambda_4
   \phi^+ \phi \sigma^3  + \lambda_5 \phi^+ \phi \sigma^2 \sigma^{*} ) \cr
                      & + ...... \cr }
} \eqno(2)$$
\no where $\phi$ is the standard model higgs doublet,
$\sigma$ siglet field with B-L number -2.
We parametrize $\sigma = {1\over \sqrt2}( \eta + V_{BL} ) e^{i \chi/ V_{BL}}$.
 This generates a non-zero mass for
the Majoron ( $\chi$ ):
  $${
m_\chi \simeq {( {25\over 2} \lambda_1 + {9\over 2} \lambda_2
        + {1\over 2}\lambda_3 )}^{1/2} {( {V_{BL} \over V_{WK}} )}^{3/2}
                      {\rm keV} ,
}\eqno(3)$$
\no for $V_{BL} > V_{WK}$. If the Majoron is stable, clearly
$V_{BL} > V_{WK}$ will be in contradiction with the cosmological mass density
constraints. The situation does not improve, for unstable Majoron, because
the only decay mode for the Majoron is to two neutrinoes. The strength
of this coupling is fixed by  low energy considerations
such as neutrino mass and is therefore quite constrained. More importantly,
it was concluded in Ref.[4] that, the upper limit on $V_{BL}$
is given by eq.(1). A sight refinement of this limit has recently been
provided in Ref.[5].

In this report, we look at the impact of non-renormalizable operators on
the Majoron decay. It is clear that
for $V_{BL} \geq 10^{8}$GeV, eq.(3) implies that,
$m_{\chi} \geq 1$TeV implying that, it could perhaps decay to a pair of Higgs
bosons (H). However, since the Higgs boson is CP-even and the Majoron is
a CP-odd particle, $\chi \rightarrow~ n~ H$ is forbidden in the limit of CP-
conservation. CP-violationg is therefore necessary for any
$\chi \rightarrow ~n~H$ decay to occur. Furthermore, no such decay can
occur from the renormalizable sector of the model. Therefore,
we must look at CP-violating non-renormalizable
Planck-induced operators. Two such operators are in eq.(2) if
$\lambda_4$ and
$\lambda_5$ are complex. They induce coupling of type:
$${
{\cal L}_{\chi}
   =  \sin \delta~~ | \lambda_5 |~~ { V_{BL}^2 \over {2 M_{Planck}} }
{}~ \chi ~ H^2 ~~,
}\eqno(4)$$
\no where we redefine the $\sigma$-field
to absorb the CP-phase of $\lambda_4$;
$\delta$ is the phase of
$\lambda_5$ in this new basis.
We limit the value of $ |\lambda_5 | \leq M_{Planck} V_{WK}^2 / V_{BL}^3$
in order that the effective Higgs mass remains of the electroweak scale. Using
this and
 $m_{\chi} \gg 2~ m_{H}$, we get for the lifetime
$\tau_{\chi}$ of the Majoron:

$${
\tau_{\chi} \simeq
 {10^{-14} \over \sin^2\delta } ~~{( {V_{BL}\over 10^{8}{\rm GeV}
                         })}^{7/2}~ sec.
}\eqno(5)$$
In order to see what constraints it implies on $V_{BL}$,
let us look at the picture of the early universe with massive Majoron
with $m_{\chi} \geq$ TeV,
which corresponds to
$V_{BL} \geq 10^{8}$ GeV.
 As was observed in Ref.[4],
below $T = m_{\nu_R} ~\simeq V_{BL}$, the
Majoron decouples and dilutes like a non-interacting gas. If its
lifetime is longer than $10^{-1}$ sec or so, it will contribute
an enormous amount to the energy density at the epoch of
nucleosynthesis {\it i.e.}
    $${
  \rho_{\chi}(T=1 {\rm MeV}) \simeq~ Y~ n_{\gamma}~ m_{\chi} ~~,
}\eqno(6)$$
\no where the factor Y is the dilution factor for the Majoron density and
  $Y \simeq {1\over 10}$[6].
This will significantly alter the predictions for Helium abundance.

On the other hand, if $\chi$ decays before
$t_U = 10^{-1} - 10^{-2} sec.$ the decay products $\chi \rightarrow
2~H \rightarrow 2b~ 2 {\overline{b}} \rightarrow 2c ~ 2{\overline{c}}
                        ~ 2l~ 2{\overline{\nu}}
\rightarrow 2d~ 2{\overline{d}}~ 4l~ 4{\overline{\nu}} $
will quickly thermalize and not have any
impact on the Helium production. Requiring
$\tau_{\chi} \leq 10^{-2} ~sec.$, we get,
    $V_{BL} \leq 10^{11}$GeV
for $\delta \simeq 0.1$. Thus,
in the presence of CP-violating Planck effects, the results of Ref.[4]
are extended as followes:
$${
V_{BL} \leq  {( { m_{\nu} \over {\rm 25 eV}} )}^{4/7} \times 10 ~{\rm TeV}
,}\eqno(7.a)$$
\no and

$${ 10^{8} {\rm GeV} \leq V_{BL} \leq 10^{11} {\rm GeV}.} \eqno(7.b)$$

A few comments are in order: there can also be heavy decay in the CP-conserving
limit such as $\chi \rightarrow t {\overline{t}},~~
b{\overline{b}}~~{\it etc}$. Their effect on these bounds is
negligible.

Another point of interest is that the decay of the massive Majoron will
generate
large amount of entropy if $\tau_{\chi} \geq 10^{-12} ~sec.$ (corresponding to
$m_{\chi} \geq 10^3 {\rm GeV}$).
This entropy generation will dilute any preexisting baryon asymmetry of the
universe by a factor $x$, which
can be obtained from energy conservation. For $m_{\chi} \gg T_{\chi}$, we find

    $$
x \equiv {S_f \over S_i} \simeq {({ {120 Y\over \pi^4 g^{*}}
     {m_{\chi} \over T_{\chi} }})}^{3/4}~~,
\eqno(8)$$
\no where $T_{\chi} \simeq {(1/ \tau_{\chi}~ {\rm in}~ sec.)}^{1/2}$MeV,
and
$g^{*}$ is the number of degrees of freedom at
$T = T_{\chi}$.
For instance, for $m_{\chi} \simeq 10^{3}$TeV,
$\tau_{\chi} \simeq 10^{-5} sec.$
leads to $x \simeq 10^{3}$.
This implies that any baryon or lepton asymmetry generated earlier than this
epoch must be of order $10^{-6} n_{\gamma}$ in order to explain present
 observation. Otherwise, the observed asymmetry must be generated
at a later time ${\it i.e}~~ t_U > \tau_{\chi}$.

In conclusion, if the non-perturbative Planck scale induced
operators are CP-violating, there appears an allowed high mass scale window
for the scale of global B-L symmetry breaking
( ${\it i.e} ~~10^{8}{\rm GeV}~ \leq~ V_{BL}~ \leq~ 10^{11}{\rm GeV}$ )
in addition to the allowed range below 10 TeV obtained in Ref.[4].

\filbreak

\bb
\b

\ce{\bf References}
\b
\item{[1]}
R. Holman, S. Hsu, T. Kephart, E. Kolb, R. Watkins and
L. Widrow, Phys. Lett. B282 (1992) 132; M. Kamionkowski and J. March-Russel,
Phys. Lett. B282 (1992) 137; S. Barr and D. Seckel, Phys. Rev. D46 (1992) 538.

\item{[2]}see, C. Misner, K. Thorne and J. Wheeler, {\it Gravitation
} (Freeman, San Francisco, 1973).

\item{[3]} Y. Chikashige, R.N. Mohapatra and R.D. Peccei,
           Phys. Lett. B98 (1980) 265.

\item{[4]}
E. Akhmedov, Z. Berezhiani, R. Mohapatra and
G. Senjanovi\'c , Phys. Lett. B299 (1993) 90.

\item{[5]}
J. Cline, K. Kainulainen and K. Olive, UMN-TH-1113-93, TPI-MINN-93/13-T,
April 1993.

\item{[6]}G. Steigman, K. Olive and D. Schramm, Phys. Rev. Lett. 43 (1979) 239;
K. Olive, D. Schramm and G. Steigman, Nucl. Phys. B180 (1981) 497.

\bye